\documentclass{article}

\usepackage{oldlfont,amssymb,epsf,stmaryrd,epsfig}

\newtheorem{proposition}{Proposition}[section]
\newtheorem{lemma}{Lemma}[section]

\newenvironment{remark}{\medskip \noindent{\bf Remark}}{}

\newcommand{\proof}[1]{{{\em Proof.} #1 $\Box$}}

\def\ra{\rightarrow}
\def\fa{\forall}
\def\ex{\exists}

\begin{document}
\title{Preliminary investigations\\
on induction over real numbers}
\author{Gilles Dowek\thanks{\'Ecole polytechnique and INRIA, 
LIX, \'Ecole polytechnique, 91128 Palaiseau Cedex, France, 
{\tt Gilles.Dowek@polytechnique.fr}}}

\date{}
\maketitle

\thispagestyle{empty}

In arithmetic, the induction principle permits to give more direct and
more intuitive proofs than alternative principles such as the
existence of a minimum to all non-empty sets of natural numbers or the
existence of a maximum to all bounded non-empty sets of natural
numbers, that usually require detours in proofs.

Moreover, when we prove a general property by induction and use it on
a particular natural number, we can eliminate the invocation of the
induction principle and get a more elementary proof \cite{Godel58}.

The idea of induction is that when a property holds for some natural
number $a$ and is hereditary, then it holds for all numbers greater
than or equal to $a$. The fact that the property is hereditary means
that when it holds for some number $n$ then it propagates ``at least
a little bit'' to greater numbers. In various formulations of the
scheme, this takes the form $P(n) \Rightarrow P(S(n))$ or $(\fa p <
n~(P~p)) \Rightarrow (P~n)$.

We present in this note a similar principle for real numbers. The fact
that some property is hereditary is expressed by the fact that if $P$
holds for a real number $c$, then its holds on an interval $[c ... c +
\varepsilon]$. Of course, this is not enough to prove that if $P$
holds for a real number $a$, then it holds for all real numbers
greater than $a$, because unlike a bounded set of natural numbers, a
bounded set of real numbers need not have a maximum. However, a closed
bounded set of real numbers does. Thus, we shall restrict our
induction principle to closed properties $P$.

In this note, we state the induction principle, we prove it and we
give several examples applications. All these examples come from
\cite{Munozetal} where we have formalized and proved results in
elementary calculus and kinematics to studying the motion of
aircraft and where this real induction principle is implicit. We also
briefly discuss which axioms would be replaced by this scheme in an
axiomatization of analysis, how this this induction scheme is related
to ordinal induction and how, in some cases, the invocation of this
scheme can be eliminated when we apply a general theorem to a
particular real number.

This work is still preliminary and several problems mentioned in this
note require further investigations.

\section{The induction scheme}

\begin{proposition}
$$[Closed(P) \wedge P(a) \wedge 
\fa c \geq a~(P(c) \Rightarrow \ex \varepsilon \fa y~(c \leq y \leq c
+ \varepsilon \Rightarrow P(y)))] \Rightarrow 
\fa x \geq a~P(x)$$
\end{proposition}

\proof{
Assume that there is a $c \geq a$ such that $\neg P(c)$ holds and consider
the set $E = \{x \in {\mathbb R}~|~x \leq c \wedge P(x)\}$. 
This set is closed, non empty and bounded hence it has a maximum $d$.
We have $P(d)$ there is an $\varepsilon > 0$ such that $P$ holds on 
$[d... d + \varepsilon]$. Let $d'$ be an element of 
$]d ... d + \varepsilon] \cap ]d ... c]$, we have 
$P(d')$ and $d' \in E$ contradicting the maximality of $d$.
}

\begin{remark} {\bf (The definition of the predicate $Closed$)}

The set $Closed$ of closed sets of real numbers can be inductively
defined as the smallest set containing the closed intervals and closed
by finite unions and intersections and by intersections of arbitrary
families.

In higher-order logic (simple type theory) with axioms stating that
the base type is a complete Archimedian totally ordered field, then
the set of real numbers have a second-order type~: $\iota \ra o$ and the
families of sets of real numbers have a third-order type~: $(\iota \ra o) \ra
o$. Hence the set $Closed$ can be defined with third-order
quantification and the induction theorem can be stated and proved in
third-order logic.

Trading definitions for axioms, we can remain in second order logic~:
we just need a third-order constant $Closed$ and axioms allowing to
build closed sets from closed intervals by finite unions and
intersections and by intersections of arbitrary families, for instance
the axiom 
$$(\fa x~Closed (\lambda y~(R~x~y))) \Rightarrow Closed(\lambda y~\fa
x~(R~x~y))$$

Using type structure, we can also define the closed sets as the
objects of type $\iota \ra o'$, where $o'$ is a sub-type of $o$ of
propositions formed with the predicates $\leq$ and $=$ and
conjunction, disjunction and universal quantification, but neither the
predicate $<$ nor negation and existential quantification.
\end{remark}

\section{Induction and the axiomatization of real numbers}

We may decide to take this induction principle as an extra axiom. In
this case, it remains to be investigated which axioms can be then
removed.

In particular the existence of a maximum for each closed bounded set
implies the completeness axiom (as if $A$ is an arbitrary non empty
set of real numbers, then $\{y \in {\mathbb R}~|~\fa x \in A~y \leq
x\}$ is a closed non empty set of real numbers and its maximum is the
least upper bound of $A$).  But it remains to be investigated if the
real induction principle implies the existence of a maximum for each
closed bounded set, or if it is weaker.

\section{An example of application}

\begin{proposition} \label{differential}
Let $a$ and $b$ be real numbers and $\alpha$ and $\beta$ two functions
from real numbers to real numbers, such that $b \geq 0$, for all $x$, $\alpha(x)
\geq 0$ and $\beta(x) > 0$. 
We consider a differentiable function $f$ solution of the differential
equation
$$f(a) = b$$
$$f'(x) = - \alpha(x) f(x) + \beta(x)$$
Then for all $x \geq a$, $f(x) \geq 0$.
\end{proposition}

\proof{We prove that if $f(x) \geq 0$, then there exists an
$\varepsilon > 0$ such that for all $h$ such that $0 \leq h \leq 
\varepsilon$, $f(x+h) \geq 0$. Let 
$$d(h) = \frac{f(x+h) - f(x)}{h} - f'(x)$$
We have 
$lim_{0}~d = 0$ and 
$$f(x+h) = f(x) + h f'(x) + h d(h) = f(x) (1 - h \alpha(x)) + h
(\beta(x) + d(h))$$ 
For $h$ small enough, we have
$1 - h \alpha(x) \geq 0$ and 
$\beta(x) + d(h) \geq 0$ (as $\beta(x) > 0$), 
hence $f(x+h)  \geq 0$.}

\begin{remark}
If we understand $f(x)$ as the position of an object in motion, then
its velocity $f'(x)$ is the sum of a negative term $- \alpha(x) f(x)$
and a positive term $\beta(x)$. The intuition is that the object
cannot cross the line $y = 0$ because when it approaches this line,
the negative term $- \alpha(x) f(x)$ vanishes and the positive term
pushes it up.

This proof where we derive constructively the theorem from the
induction principle and where we investigate the motion of the object
in a near future and balance the action of the negative and positive
terms reflects this intuition much better than the following
alternative proof, using a reduction {\em ab absurdum}.

\proof{Assume there is some $c$ such that $f(c) < 0$. Then call $b =
max \{x \leq c~|~f(x) = 0\}$ the last time the function crosses the
line $y = 0$ before $c$. Using Rolle's theorem, there is a point $t$, 
$b \leq t \leq c$ such that 
$$f'(t) = \frac{f(c) - f(b)}{c - b} < 0$$
As $f$ does not cross the line $y = 0$ between $b$ and $c$ we have
$f(t) < 0$ and thus 
$$f'(t) = - \alpha(t) f(t) + \beta(t) \geq 0$$
yielding a contraction.}

Notice, however, that this indirect proof allows to prove the theorem
with the hypothesis $\beta(x) \geq 0$, while our proof needs
the stronger hypothesis $\beta(x) > 0$.  It remains to be investigated if we
can constructively derive this theorem with the hypothesis $\beta(x)
\geq 0$ from either Rolle's theorem or the real induction principle.
\end{remark}

\section{Real induction and ordinal induction}

Let $P$ be a closed property and $a$ a real number such that 
$$P(a)$$
$$\fa c \geq a~(P(c) \Rightarrow \ex \varepsilon \fa y~(c \leq y \leq c
+ \varepsilon \Rightarrow P(y)))$$

We skolemize this hypothesis assume that we have a function $f$
mapping every real number $c$ to a real number $c' > c$ such that if
$P$ holds on 
$c$ then it holds on the interval $[c ... c']$.

Consider the function $F$ from ordinals to ${\mathbb R}$ defined as
follows. 
\begin{itemize}
\item $F(0) = a$
\item $F(S(x)) = f(F(x))$
\item if $x$ is a limit ordinal and $\{F(y)~|~y < x\}$ is bounded then 
$F(x) = sup \{F(y)~|~y < x\}$. 
Otherwise $F(x)$ is not defined. 
\end{itemize}

Let $\alpha$ be the collection of ordinal on which $F$ is defined.
The collection $\alpha$ is either the collection of ordinals or an
ordinal (in this case it can only be a limit ordinal).  It cannot be
the collection of ordinals, because otherwise we could associate a
real number to any ordinal, in particular to any ordinal of
$2^{2^{\aleph_{0}}}$ and thus we would have an injection from
$2^{2^{\aleph_{0}}}$ to ${\mathbb R}$ which is impossible for
cardinality reasons. Thus $\alpha$ is an ordinal.

We can assign this way an ordinal to the pair $(a,f)$, measuring the
complexity of this proof by induction. 

As there is always a rational number between $F(x)$ and $F(S(x))$,
this ordinal is denumerable.  Alexandre Miquel has proved that,
conversely, any denumerable ordinal can be embedded into ${\mathbb R}$
\cite{Miquel}. Thus, for every denumerable ordinal $\alpha$, there is
a pair $(a,f)$ to which $\alpha$ is associated.

Notice that an ordinal can be embedded in ${\mathbb R}$ if and only if
it is denumerable, and this whether or not the continuum hypothesis
holds.

Thus, real induction seems to be equivalent to $\aleph_{1}$-induction.

\section{Proof reduction}

In arithmetic, when we prove a theorem $\fa x~P(x)$ by induction,
using a proof $P(0)$ and a proof of $\fa x~(P(x) \Rightarrow P(S(x)))$
and we apply this proof to an explicitly given natural number $n$, we
can transform this proof into an elementary proof not using the
induction scheme, proving successively $P(1)$, $P(2)$, $P(3)$
... $P(n)$.

With real induction this is sufficient if the monotonous sequence
$f^n(a)$ is unbounded.  But this sequence may also be bounded and
hence convergent and in this case applying the induction hypothesis a
finite number of times is not sufficient.

For instance if $f(0) = 1/2$ and $f(1-1/2^n) = 1-1/2^{n+1}$. From
$P(0)$ we obtain that $P$ holds on the interval $[0 ... 1/2]$, then on
the interval $[1/2 ... 3/4]$, then on $[3/4 ... 7/8]$, $[7/8
...15/16]$... never reaching the value $1$. 

To prove $P(1)$, we have to use the fact that as $P$ is a closed set
and the sequence $1-1/2^{n+1}$ is in $P$, its limit $1$ is also in
$P$.

Thus, to get an ``elementary'' proof of $P(1)$ we must first build a
proof of $\fa n~P(f^{n}(0))$ using recursion on natural numbers, and
then a proof of $P(1)$ using this proof and a proof that the sequence
$f^{n}(0)$ has $1$ as limit and that $P$ is closed.

Then starting again from $1$ we may apply the induction hypothesis and
either the sequence $f^{n}(1)$ is unbounded or convergent and, in this
case, we must take the limit again...

Notice that to construct this elementary proof, we need either to know
that the monotonous sequence 
$f^{n}(0)$ has a limit $l$ it is unbounded. In other words, we need to
have a constructive proof of the proposition 
$$\ex l~(Lim~\lambda n~f^{n}(0)~l) \vee \fa x~\ex n f^{n}(0) \geq x$$

Thus we need an ordinal $\alpha$, and a function $F$ from $\alpha$ to
${\mathbb R}$ together with proofs that
\begin{itemize}
\item $F(0) = a$,
\item $F(S(x)) = f(F(x))$,
\item if $x$ is a limit ordinal $< \alpha$ then 
$\{F(y)~|~y < x\}$ is bounded and
$F(x) = sup \{F(y)~|~y < x\}$. 
\item $\{F(y)~|~y < \alpha\}$ is unbounded.
\end{itemize}

\section{An application to a kinematics problem}

We consider a point in motion whose trajectory is a function $M(t) =
(x(t), y(t))$.  We choose the origin of the coordinate system at
$M(0)$. Thus $x(0) = y(0) = 0$.  The velocity vector of this point is
$$M'(t) = (x'(t), y'(t)) = (v(t) cos (\theta(t)), v(t) sin
(\theta(t)))$$ We assume that the intensity of the velocity vector is
constant $v(t) = v$ and that the derivative of its polar angle is
bounded $-\rho \leq \theta'(t) \leq \rho$.  Such an hypothesis is
realistic when studying the motion of aircraft because the derivative
of the heading of an aircraft depends on its bank angle that is
bounded for obvious reasons.

We consider the position of the point at time $t$ and we want to
evaluate its minimum distance to the origin.

Let us consider two situations. First, if at all times $t$,
$\theta'(t) = 0$ the motion of the point is a straight line and the
distance to the origin at time $t$ is $v t$. Second, if at all times
$t$, $\theta'(t) = \rho$ the motion is a circle $x(t) = (v / \rho)
(cos (\rho t) - 1)$, $y(t) = (v / \rho) sin (\rho t)$ and the distance
to the origin at time $t$ is $F(t) = (2v/\rho) |sin (\rho t/2)|$.

\begin{center}
\epsfig{file=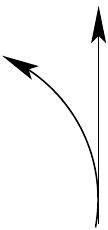}
\end{center}

We conjecture that for a time $t < 2 \pi / \rho$ the distance 
$F(t) = (v/\rho) sin (\rho t/2)$ is the minimum distance of the point to
the origin (for $t \geq 2 \pi / \rho$ the minimum distance is obviously $0$
as the point can come back to the origin at time $t$ after a circle of
appropriate radius). 

We prove the result for $t < 2 / \rho$. 

Consider the position of the point $x(t) = R(t) cos(\alpha(t))$, $y(t)
= R(t) sin(\theta(t)$. 
We first prove the following lemma 

\begin{lemma}
If for $0 < t < x$,
$R'(t) > 0$ then, on this same interval,
$-\rho /2 \leq \alpha'(t) < \rho / 2$. 
\end{lemma}

\proof{
We have 
$$x'(t) = R'(t) cos (\alpha(t)) - R(t) sin (\alpha(t)) \alpha'(t) = v
cos(\theta(t))$$
$$y'(t) = R'(t) sin(\alpha(t)) + R(t) cos (\alpha(t)) \alpha'(t) = v
sin(\theta(t))$$
Thus
$$R'(t) = v (cos(\theta(t)) cos (\alpha(t)) + sin(\theta(t) sin
(\alpha(t)))) = v cos (\theta(t) - \alpha(t))$$
$$R(t) \alpha'(t) = v (sin(\theta(t)) cos (\alpha(t)) - cos(\theta(t) sin
(\alpha(t)))) = v sin (\theta(t) - \alpha(t))$$
$$(R(t) \alpha'(t))' = 
R'(t) \alpha'(t) + R(t) \alpha''(t) = v cos (\theta(t) - \alpha(t))
(\theta'(t) - \alpha'(t)) = R'(t) (\theta'(t) - \alpha'(t))$$
Thus 
$$\alpha''(t) = R'(t)/R(t) (\theta'(t) - 2 \alpha'(t))$$
Let $\varepsilon > 0$, consider the function 
$f(t) = \alpha'(t) + \rho/2 + \varepsilon$
the differential equation above rewrites 
$$f'(t) = -2 R'(t)/R(t) f(t) + R'(t)/R(t) (\theta'(t) + \rho + 2
\varepsilon)$$ 
We have $R'(t)/R(t) \geq 0$ and 
$R'(t)/R(t) (\theta'(t) + \rho + 2 \varepsilon) > 0$. Hence 
by proposition \ref{differential}, we get $f(t) \geq 0$.

Thus for all $\varepsilon > 0$, 
$\alpha'(t) + \rho/2 + \varepsilon \geq 0$
hence 
$$- \rho/2 \leq \alpha'(t)$$
By a similar argument (considering $g(t) = - \alpha'(t) + \rho/2 +
\varepsilon$)
we obtain
$$\alpha'(t) \leq \rho/2$$
}

We then show that we can get rid of the hypothesis $R'(t) > 0$,
i.e. that  for at any time $x < 2/\rho$, we have $-\rho/2 \leq
\alpha'(x) \leq \rho/2$.

\begin{lemma}
If $x < 2/\rho$ then $-\rho/2 \leq \alpha'(x) \leq \rho/2$.
\end{lemma}

\proof{ We use the induction schema and consider the (closed property)
$|\alpha'(x)| \leq \rho/2$ and $|R'(x)| \geq 0$.  We assume that the
property holds for all times $t \leq x$ and we prove that it holds a
little bit more.

At all times $t \leq x < 2/\rho $ we have $R(t) \leq vt <
2v/\rho$. Thus $R(t) \alpha'(t) \leq vt < v$.  As $(R(t) \alpha'(t))^2
+ (R'(t))^2 = v^2$ we get $R'(t) \neq 0$ and hence $R'(t) > 0$. Thus
we have $R'(t) > 0$ on an interval $[x ...  x + \varepsilon]$ for some
$\varepsilon$. By the previous lemma, on this interval $|\alpha'(x)|
\leq \rho/2$ and $|R'(x)| \geq 0$.  }

Then we are ready to conclude 
\begin{proposition}
For $t < 2/\rho$, $R(t) \geq F(t)$.
\end{proposition}
\proof{
Let $\varepsilon > 0$, $\rho' = \rho + \varepsilon$ and
$F_{\varepsilon}(t) = 2v/\rho' |sin (\rho' t/2)|$. 

We have $F_{\varepsilon}'(t) = v/\rho' cos (\rho' t/2)$ and thus 
$$F_{\varepsilon}'^2(t) + (\rho'/2)^2 F_{\varepsilon}^2(t) = v^2$$
and we already know that
$$R'^2(t) + \alpha'^2(t) R^2(t) = v^2$$

Thus 
$$R'^2(t) - F_{\varepsilon}^2(t) = - \alpha'^2(t) R^2(t) +
(\rho'/2)^2 F_{\varepsilon}^2(t)$$ 
$$(R'(t) - F_{\varepsilon}'(t))(R'(t) + F_{\varepsilon}'(t)) = - \alpha'^2(t) (R(t)-F_{\varepsilon}(t))(R(t) +
F_{\varepsilon}(t)) + ((\rho'/2)^2 - \alpha'(t)^2) F_{\varepsilon}^2(t)$$
Call $S$ the function $R - F_{\varepsilon}$
We have 
$$S'(t) = - [\alpha'^2(t) (R(t) + F_{\varepsilon}(t)) / (R'(t) +
F_{\varepsilon}'(t))] S(t) + [((\rho'/2)^2 - \alpha'(t)^2)
F_{\varepsilon}^2(t) (R'(t) + F_{\varepsilon}'(t)) / (R'(t) +
F_{\varepsilon}'(t))]$$ We have $R' \geq 0$, $(\rho'/2)^2 >
\alpha'(t)^2$ and $F_{\varepsilon}^2(t) > 0$ thus applying again
proposition \ref{differential}, we obtain $S(t) \geq 0$. Hence $R(t)
\geq F_{\varepsilon} (t)$.

We have $R(t) \geq F_{\varepsilon} (t)$ for all $\varepsilon > 0$, thus 
$R(t) \geq F(t)$ 
}


\begin{thebibliography}{99.}

\bibitem{Godel58}

K. G\"{o}del, 
\"{U}ber eine bisher noch nicht ben\"{u}tzte Erweiterung des finiten
Standpunktes, 
{\em Dialectica}, 12, pp. 280-287.

\bibitem{Miquel}
A. Miquel, Embedding ordinals into the real line, {\em manuscript}
(2002). 

\bibitem{Munozetal}
C. Mu\~{n}oz, R. Butler, V. Carre\~{n}o and G. Dowek, Formal
verification of conflict detection algorithms, 
{\em Conference on Correct
Hardware Design and 
Verification Methods}, Lecture Notes in Computer Science 2144,
Springer-Verlag (2001), pp. 403-417. Technical report,
NASA/TM-2001-210864. 


\end{thebibliography}
\end{document}